# The feasibility of Q-band millimeter wave on hand-gesture recognition for indoor FTTR scenario

Yuxuan Hu, Zhaoyang Xia, Yanbo Zhao, Feng Xu
Key lab of Electromagnetic Wave Information Science, Fudan University
200433, Shanghai, China

*Abstract*—The generalization for different scenarios and different users is an urgent problem for millimeter wave gesture recognition for indoor fiber-to-the-room (FTTR) scenario. In order to solve this problem and verify the feasibility of FTTR Q-band millimeter wave in gesture recognition, we build a real-time millimeter wave gesture recognition system. The moving hand-gestures are represented as a variety of time-variant spectrum features, such as micro-Doppler feature, and then the feature learning and classification is realized by using a convolution neural network (CNN). The experimental results show that the millimeter wave gesture recognition system can achieve the generalized gesture recognition for 2 scenarios and 4 users.

## I. Introduction

The traditional indoor Wi-Fi has been unable to meet the increasing network demand of people due to limited transmission rate of network cable, small bandwidth, and serious quality degradation after passing through the wall [1]. Fiber-to-the-Room (FTTR) will become the common network scheme in the future by laying optical fiber to every room to achieve high-quality network [1]. The Q-band millimeter wave can match FTTR best due to its wide available bandwidth and non-interference across rooms. With the extensive deployment of Q-band FTTR, millimeter wave communication and sensing will become very promising technologies.

This paper mainly explores the feasibility of Q-band millimeter wave in hand-gesture recognition for indoor FTTR scenario. The millimeter wave gesture recognition faces the challenges of long distance and transfer learning for user and scenario for indoor FTTR scenario, as shown in Fig. 1.

The hand-gesture recognition method using non-wideband electromagnetic wave such as Wi-Fi [2] has poor anti-interference ability and weak range resolution, which cannot solve these challenges. In contrast, the gesture recognition method using wideband millimeter wave can achieve high range resolution, and has the advantages of anti-interference, small antenna size and sensing submillimeter level micro-motion [3].

At present, the research of millimeter wave gesture recognition has made some progress. In 2015, Google demonstrated the Soli [3] [4] project of using 60GHz band millimeter wave to achieve hand-gesture recognition in close range, and the application value of millimeter wave hand-gesture recognition began to attract great interest. Lien et al. [3] adopted the random forest method with more than 700 features extracted from echo of the hand-gestures, which can achieve an average classification accuracy of 92.1% for 4 classes of hand-gestures. Wang et al. [4] exploited the end-to-end convolution neural network (CNN) method with range-doppler (RD) feature, which can achieve an average classification accuracy of 87% for 11 classes of hand-gestures. Hazra et al. [5] used the 60GHz band millimeter wave to obtain the range-Doppler image sequence, and input a long recurrent all-convolution neural network, which can achieve the average classification accuracy of 94.34% for 5 classes of hand-gestures. Zhang et al. [6] obtained the range spectrum feature using the 24GHz band millimeter wave, and based on a recurrent three-dimensional (3D) CNN+LSTM method, which can achieve an average classification accuracy of 96% for 8 classes of hand-gestures. However, these studies do not involve the generalization of scenarios, locations and users.

Xia et al. [7] studied the difference of recognition performance for different users. First, the two-dimensional (2D) time spectrum features representing the variations of range, Doppler, azimuth and elevation are obtained by using the 77GHz band millimeter wave. And a multi-channel CNN is designed to classify 8 classes of hand-gestures. The average classification accuracies of training user A, non-training user B and non-training user C are 98.9%, 79.8% and 89.3% respectively. Xia et al. [8] also studied the impact of different locations and different users on recognition performance. Firstly, the moving point cloud of hand-gestures in the range of 1m ~ 3m, - 30 ° ~ 30 ° is obtained by using the 60GHz band millimeter wave. And the time spectrum features representing the 3D spatial position variations of the hand-gestures are extracted from the time-varying moving point cloud. The 3 time spectrum are input into multi-channel CNN to classify 5 classes of gestures. The average classification accuracies of training user A, non-training user B and non-training user C are 99.7%, 91.1% and 87.1% respectively.

This paper studies the influence of user and scenario on recognition performance, and is committed to achieving hand-gesture recognition that generalizes to different users and different scenarios.

## II. GESTURE FEATURES

The commercial Multiple Input Multiple Output (MIMO) millimeter wave radar can distinguish the moving target into multiple target points with complex amplitude in the range Doppler domain. And the target point cloud can be obtained by using the angle estimation method to further calculate the azimuth and elevation of each target point. According to the moving scattering center model in [7], the moving target can be represented as $N_p$ moving scattering centers or a point cloud set [8] with the parameters of range $r_p$, radial velocity $v_p$, azimuth $\theta_p$, elevation $\varphi_p$ and complex scattering amplitude $\widetilde{A_p}$.

Using the point cloud dimension reduction method in [8], the time-variant point cloud of motion gesture can be represented as four time-variant 2D features, including range-time-amplitude (RTA) feature, azimuth-time-amplitude (ATA) feature, elevation-time-amplitude (ETA) feature and Doppler-time-amplitude (DTA) feature. In order to fully represent the 3D motion hand-gestures, the radial feature RTA or DTA, azimuth feature ATA and elevation feature ATA can be combined. In this paper, DTA, ATA and VTA are combined to represent the motion hand-gestures.

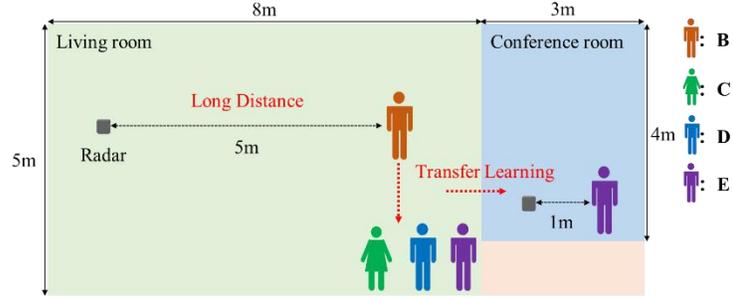

Fig. 1. The challenges of millimeter wave gesture recognition for indoor FTTR scenario.

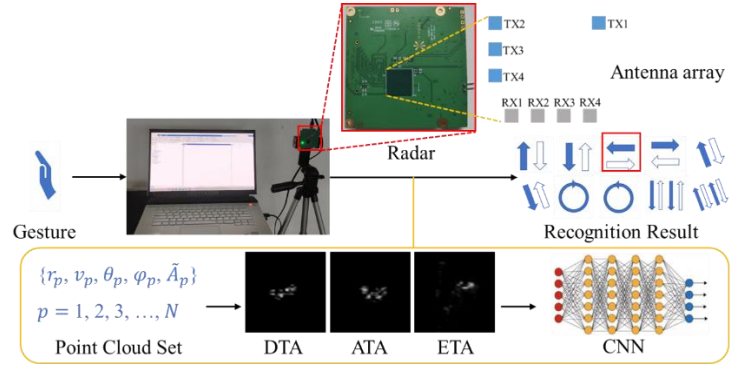

Fig. 2. Real-time millimeter wave gesture recognition system

## III. EXPERIMENTS AND ANALYSES

### A. Real-Time Millimeter Wave System

In order to verify the feasibility of millimeter wave gesture recognition in the FTTR scenario, we design a real-time millimeter wave gesture recognition system as shown in Fig. 2. The device for transmitting and receiving millimeter wave is the CAL60S244-IBM-AiP millimeter wave radar evaluation board of Calterah. The device consists of four transmit antennas and four receive antennas which forms a 2D virtual antenna array in a time-division multiplexed (TDM) MIMO mode. The sweep frequency range is set to 60.5 ~ 64GHz, the frame rate is 20 frames/s, the number of frequency sweep cycles per frame is 256, and the number of sampling points per frequency sweep cycle is 256.

### B. Gesture Data set

We designed 10 classes of hand-gestures, including wave up, wave down, wave left, wave right, push, pull, circle clockwise, circle anticlockwise, double-click, and double-push, as shown in Fig. 3. In order to obtain the three 2D gesture features of DTA, ATA and ETA, a sliding window with a length of 30 frames is used to capture the motion hand-gesture. The capture condition is set to the maximum radial velocity greater than 0.3 m/s and the number of motion frames greater than 5. The three features of 10 classes of hand-gestures at the gesture interaction position are shown in Fig. 3. The gesture interaction position is around the azimuth of 0 ° at the distance of 1 m relative to radar.

In order to test the gesture recognition performance of different scenarios and different users, it is necessary to establish a dataset containing multiple scenarios and multiple users at the gesture interaction position.

First, in a living room scenario, the training user B, training user C, and training user D perform each class of hand-gesture 40-60 times to obtain the training data set.

Then the training user B, training user C, training user D, and non-training user E perform each class of gesture in the living room scenario and the conference room scenario for about 30 times respectively to obtain the gesture data sets of two scenarios and four users. The test user does not participate in the training.

## C. Gesture classification

The training data set is divided into training set and verification set at a ratio of 7:3, and the lightweight CNN in [8] is used for training to obtain the gesture classification model. The number of channels in the input layer is set according to the number of the input features, including DTA, ATA, and ETA. And the adam optimizer is used to iterate 100 Epochs with a constant learning rate of 0.001. The number of batch samples in each iteration is 32.

The real-time recognition results of 2 scenarios and 4 users by the millimeter wave gesture recognition system with the gesture classification model are shown in Table I. It can be seen that the average classification accuracy of the non-training user is only slightly lower than that of the worst performing training user in the living room and conference room scenarios. In addition, there is no significant difference in gesture recognition performance between the two scenarios. As shown in Table II, the recognition results of the Inception-v3 for the same scenes and gesture data indicate that the scene generalization and user generalization capabilities of the Inception-v3 are weaker than those of the lightweight CNN. It shows that the gesture recognition system can realize the generalization for users and scenarios.

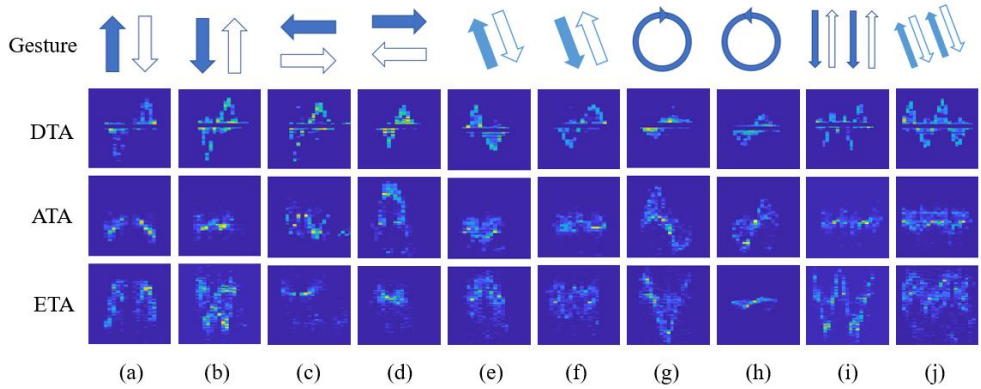

Fig. 3. Gesture diagrams and feature images for (a) Wave up, (b) Wave down, (c) Wave left, (d) Wave right, (e) Push, (f) Pull, (g) Circle clockwise, (h) Circle anticlockwise, (i) Double-click and (j) Double-push.

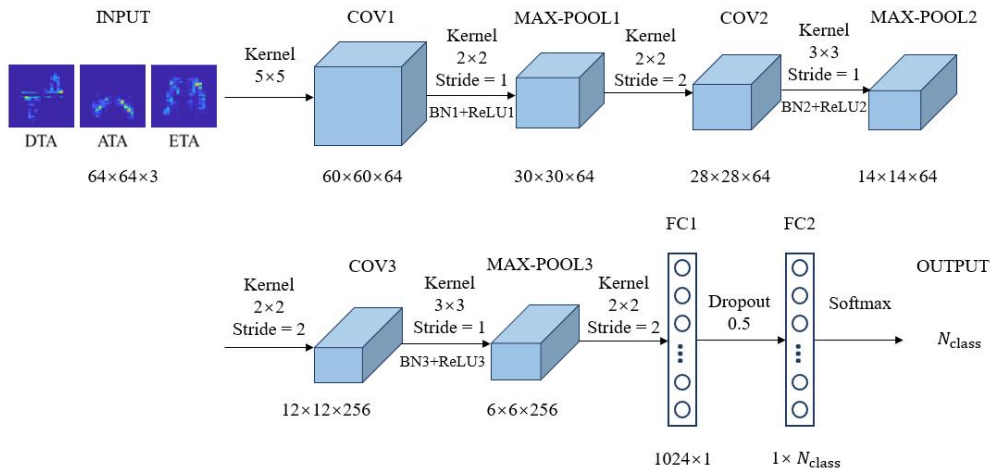

Fig. 4. Lightweight CNN structure.

TABLE I
REAL-TIME RECOGNITION RESULTS OF MILLIMETER WAVE GESTURE RECOGNITION SYSTEM (*INCEPTION-V3*)

| Scene | Average accuracy | | | |
|---|---|---|---|---|
| | Training User B | Training User C | Training User D | Non-training User E |
| Living room | 98.6% | 97.9% | 97.4% | 85.7% |
| Conference room | 86.0% | 97.8% | 94.1% | 80.8% |

TABLE II
REAL-TIME RECOGNITION RESULTS OF MILLIMETER WAVE GESTURE RECOGNITION SYSTEM (LIGHTWEIGHT CNN)

| Scene | Average accuracy | | | |
|---|---|---|---|---|
| | Training User B | Training User C | Training User D | Non-training User E |
| Living room | 97.5% | 98.7% | 95.7% | 95.4% |
| Conference room | 96.0% | 98.2% | 95.7% | 93.8% |

## IV. CONCLUSION

The real-time millimeter wave gesture recognition system designed in this paper can realize the generalized recognition when transferring from training scenario and training user to non-training scenario and non-training user. It proves the feasibility of millimeter wave gesture recognition applied to indoor FTTR scenario. However, for the real Q-band FTTR scenario, the modulation mode of millimeter wave is determined by the communication requirements, rather than the linear frequency modulation adopted in this paper. In addition, the signal-to-noise ratio and the number of point cloud of moving hand-gestures will decrease with the increase of the millimeter wave signal propagation distance, which poses a challenge for gesture recognition at a relatively long distance. In order to further improve the practicability of millimeter wave gesture recognition in FTTR scenario, we will focus on the optimization of sparse point cloud at a long distance and the integrated sensing and communication (ISAC) technology using communication signals for sensing in the future.